\documentclass[reprint,superscriptaddress,amsmath,amssymb,prl,floatfix,footinbib]{revtex4-1}

\usepackage[pdftex]{graphicx}
\usepackage{bm}
\usepackage{color}
\usepackage{times}
\usepackage{amsmath}
\usepackage{amssymb}
\usepackage{mathtools}
\usepackage{soul}
\usepackage{cleveref}
\usepackage{enumitem}

\newcommand\underrel[3][]{\mathrel{\mathop{#3}\limits_{%
      \ifx c#1\relax\mathclap{#2}\else#2\fi}}}

\begin{document}

\title{Dynamics of  Run-and-Tumble Particles in Dense Single-File Systems}

\author{Thibault Bertrand}
\email[Electronic address: ]{thibault.bertrand@upmc.fr}
\affiliation{Laboratoire Jean Perrin, UMR 8237 CNRS, Sorbonne Universit\'{e}, 75005 Paris, France}
\author{Pierre Illien}
\affiliation{EC2M, CNRS UMR7083 Gulliver, ESPCI Paris, PSL Research University, 75005 Paris, France}
\author{Olivier B\'{e}nichou}
\affiliation{Laboratoire de Physique Th\'{e}orique de la Mati\`{e}re Condens\'{e}e, UMR 7600 CNRS, Sorbonne Universit\'{e}, 75005 Paris, France}
\author{Rapha\"{e}l Voituriez}
\email[Electronic address: ]{voiturie@lptmc.jussieu.fr}
\affiliation{Laboratoire Jean Perrin, UMR 8237 CNRS, Sorbonne Universit\'{e}, 75005 Paris, France}
\affiliation{Laboratoire de Physique Th\'{e}orique de la Mati\`{e}re Condens\'{e}e, UMR 7600 CNRS, Sorbonne Universit\'{e}, 75005 Paris, France}

\date{\today}


\begin{abstract}
We study a minimal model of  self-propelled particle in a crowded single-file environment. We extend classical models of exclusion processes (previously analyzed for diffusive and driven tracer particles) to the case where the tracer particle is a run-and-tumble particle (RTP), while all bath particles perform symmetric random walks. In the limit of high density of bath particles, we derive exact expressions for the full distribution $\mathcal{P}_n(X)$ of the RTP position $X$ and all its cumulants, valid for arbitrary values of the tumbling probability $\alpha$ and  time $n$. Our results highlight striking effects of crowding on the dynamics: even cumulants of the RTP position are increasing functions of $\alpha$ at intermediate timescales, and display a subdiffusive anomalous scaling $\propto \sqrt{n}$ independent of $\alpha$ in the limit of long times $n\to \infty$. These analytical results set the ground for a quantitative analysis of experimental trajectories of real biological or artificial microswimmers in extreme confinement. 
\end{abstract}

\maketitle


Stemming from experimental observations of bacterial motion \citep{berg-book-2004}, Run-and-tumble particles (RTPs) provide a canonical model for the theoretical description of biological or artificial self-propelled entities such as janus particles, bacteria \citep{berg-book-2004,schnitzer-pre-1993,dileonardo-pnas-2010,saragosti-pnas-2011,schwarz-linek-pnas-2012}, algae \citep{polin-science-2009}, eukaryotic cells \cite{heuze-immunolrev-2013}, or larger scale animals \cite{benichou-rmp-2011}. In these examples of so-called active particles,  self-propulsion results from the conversion  of energy supplied by the environment into mechanical work   \citep{ramaswamy-arcmp-2010,bechinger-rmp-2016}. In its simplest form, RTP trajectories consists of a sequence of randomly oriented 'runs' --- periods of persistent motion in straight line at constant speed --- interrupted by  instantaneous changes of direction (also named polarity of the RTP), called 'tumbles', occurring at random with constant rate. 

The interplay between active particles and their environment has attracted significant interest recently~\cite{bechinger-rmp-2016}. Indeed, most motile biological systems such as bacteria or mammalian cells navigate
disordered and complex natural environments such as soils, soft gels ({\it e.g.} mucus or agar) or tissues. Through their interactions with  the environment, RTPs display robust non-equilibrium features, focus of many works. For example, recent simulations explored the dynamics of active particles in the presence of quenched disorder as well as active baths \citep{pince-natcomm-2016,volpe-pnas-2017,zeitz-epje-2017,sandor-pre-2017,reichhardt-pre-2015}. In confined geometries, active particles were found to accumulate at the boundaries, at odds with the equilibrium Boltzmann distribution~\citep{berke-prl-2008,wensink-pre-2008,elgeti-epl-2009,elgeti-epl-2013,vladescu-prl-2014,bricard-natcomm-2015}. Such non trivial interactions with obstacles can lead to effective trapping and thus have important consequences in the dynamics of self-propelled particles in disordered environments~\citep{chepizhko-prl-2013b}. For example, it was recently shown that the large scale diffusivity of RTPs moving in a dynamic crowded environment is nonmonotonic in the tumbling rate for low enough obstacle mobility in dimension $d\ge 2$ \citep{bertrand-prl-2018}.

Effects of crowding are known to have particularly strong  consequences in one-dimensional systems. In the classical example of single-file diffusion,  where identical passive particles diffuse on a line,  hard-core interactions impose conservation of the ordering of particles, thereby inducing long lived correlations in the motion of a tagged particle and eventually a subdiffusive scaling of its mean squared displacement, $\mathrm{MSD} \propto \sqrt{n}$ with time $n$ \citep{harris-jap-1965,levitt-pra-1973,fedders-prb-1978,alexander-prb-1978,arratia-ap-1983,lizana-pre-2010,taloni-pre-2008,taloni-softmatter-2017}. Physically, this results from the fact that displacements on increasingly large distances need to mobilize the motion of an increasingly large number of particles. Single file diffusion has been experimentally observed in a variety of natural and man-made materials ranging from passive rheology in zeolites \citep{karger-book-1992,hahn-prl-1996}, transport of colloidal particles under confinement \citep{wei-science-2000,lutz-jpcm-2004,lin-prl-2005,chou-jchemphys-2006,locatelli-prl-2016}, to diffusion of water in carbon nanotubes \citep{striolo-nanoletters-2006,cambre-prl-2010}. 

Diffusion under extreme confinement is also seen in biological settings with examples ranging from DNA translocation \citep{kasianowicz-pnas-1996,meller-prl-2001}, transport of proteins in crowded fluids like the cytoplasm \citep{bressloff-rmp-2013}, transport of ions in membrane channels \citep{lea-jtheobiol-1963,jensen-pnas-2010} and even migration of dendritic cells in lymphatic vessels \citep{heuze-immunolrev-2013}. Theoretically, recent works have studied the dynamics of active and biased tracer particles in single-file systems \citep{hanggi-rmp-2009,locatelli-pre-2015,cividini-jsmte-2016,kundu-epl-2016}. Exact predictions for the full distribution of particle positions were derived in the case of a tracer particle driven out of equilibrium by external forcing in a dense single-file environment \citep{illien-prl-2013}. Despite few effort, analytical results for the dynamics of self-propelled particles in complex and confined environments are still largely missing. 
 
In this Letter, we study a minimal model of self-propelled particle in a crowded single-file environment. Despite the challenge of the inherent coupling between the dynamic environment of the tracer and its polarity, we derive exact analytical results for the dynamics of RTP in the limit of high density of diffusive bath particles; in particular, we provide expressions for the full distribution $\mathcal{P}_n(X)$ of the RTP position $X$ and all its cumulants, valid for arbitrary values of the tumbling probability $\alpha$ and  time $n$. Our results highlight striking effects of crowding on the dynamics. We show in particular that even cumulants of the RTP position are increasing functions of $\alpha$ at intermediate timescales, and display a subdiffusive anomalous scaling $\propto \sqrt{n}$ with a prefactor {\it independent} of $\alpha$ in the limit $n\to \infty$. We show a perfect agreement between our analytical predictions and the results of numerical simulations. We generalize to RTPs questions that have attracted attention for passively diffusing and externally driven tracers. 

{\it Model and RTP polarity dynamics ---} In what follows, we consider a model discrete in space and in time. We take a one-dimensional lattice, infinite in both directions. The lattice sites are occupied by particles with mean density $\rho$ performing symmetric random walks; the bath particles interact via hard-core repulsion, {\it i.e.} the occupancy number of each lattice site is at most equal to one. At time $n=0$, we place at the origin a RTP. In absence of bath particles, the RTP moves along the direction of its polarity with probability 1. Thus, if its polarity is along $\bm{e}_1$ (respectively, $\bm{e}_{-1}$), the RTP moves to its right with probability $p_1 = 1$ and to its left with probability $p_{-1}=0$ (respectively, $p_1 = 0$ and  $p_{-1} =1$). Further, the polarity direction is reversed at each timestep with {\it tumbling probability} $\alpha$ (see {\it Supplemental Material} \footnote{Supplemental Material available online at ...} for details). As opposed to externally driven tracers previously studied \citep{illien-prl-2013,benichou-prl-2014}, we treat here the case of self-propelled particles. Our model then corresponds to the limit of a self-propelled particle with infinite P\'eclet number \citep{bechinger-rmp-2016}.

\begin{figure}[h]
\centering
\includegraphics[width=0.45\textwidth]{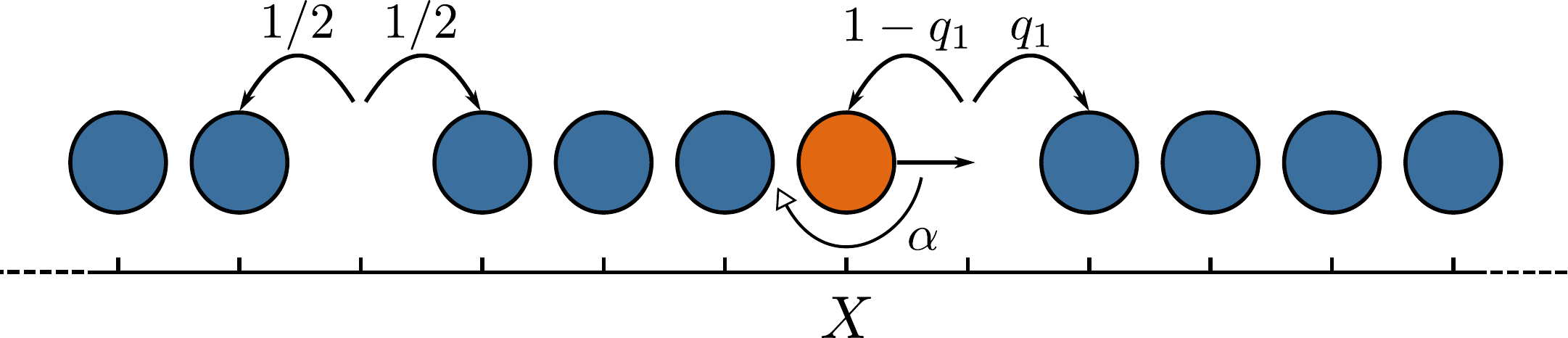}
\caption{{\it Model and Evolution Rules} - The RTP (orange) moves through a bath of diffusive particles (blue), its polarity flips at each time step with probability $\alpha$; the RTP moves by exchanging positions with vacancies.}
\label{fig:drawing}
\end{figure}

We focus here on the limit of dense systems, {\it i.e.} the limit of low vacancy densities $\rho_0 = 1-\rho \ll 1$. Following \citep{brummelhuis-physa-1989}, we formulate in this limit the dynamics of the vacancies rather than that of the particles which equivalently encodes the dynamics of the whole system. We follow the evolution rules previously established by \citep{brummelhuis-jstatphys-1988,brummelhuis-physa-1989,benichou-pre-2002,illien-prl-2013} in which we assume that each vacancy performs a nearest-neighbor symmetric random walk everywhere on the lattice except in the vicinity of the RTP. When surrounded by bath particles, a vacancy thus moves to one of its neighboring sites with equal probability. However, when the RTP lies on one of its adjacent sites, we accomodate the nature of the RTP dynamics by implementing the following specific rules (as detailed in \citep{Note1}): if the vacancy occupies the site to the right (respectively, to the left) of the RTP, it has a probability $q_1 = 1/(2p_1+1)$ (respectively, $q_{-1} = 1/(2p_{-1}+1)$ to jump to the right (respectively, to the left) and $1-q_1$ (respectively, $1-q_{-1}$) to jump to the left (respectively, to the right). It is important to note that additional rules for the cases where two vacancies are adjacent or have common neighbors would normally be needed to complete the description of the dynamics; however, these cases lead to corrections in $\mathcal{O}(\rho_0^2)$ and are thus unnecessary here. Although these hopping probabilities implicitly depend on the RTP polarity at a given time, the decoupling between the dynamics of the polarity and the vacancies makes analytical calculation tractable.

{\it Systems with a single vacancy ---} Following \citet{brummelhuis-jstatphys-1988}, we first consider an auxiliary problem: a system containing a single vacancy. Trivially in the single vacancy case, the RTP can only be in one of two positions which depend on the original location of the vacancy: if the vacancy starts in $Z>0$ (respectively, $Z<0$), the RTP can be found in $X=0$ or $X=1$ (respectively, in $X=-1$ or $X=0$). The dynamics of the RTP is dictated by exchanges of positions with the vacancies, as such it is controlled by the first-passage statistics of vacancies to the position of the RTP. Clearly, there are four cases to consider in general, but symmetry considerations reduce those cases to two: for a given initial RTP polarity, the vacancy can start either on the side of the RTP the polarity points to or not. 

We write $p_{\nu}^{(n)} (X|Z)$ the probability to find the RTP in position $X$ at time $n$, knowing that it started in $X=0$ with polarity $\nu$ and the vacancy at position $Z$. We can represent  this quantity as a sum over all passage events of the vacancy to the RTP location through the following recurrence relation \citep{brummelhuis-physa-1989,benichou-pre-2002,illien-prl-2013}
\begin{align}
p_{\nu}^{(n)}&(X|Z) = \delta_{X,0}\left( 1-\sum_{k=0}^{n} \mathcal{M}_{\nu}^{(k)}(Z) \right) \nonumber \\ 
&+ \sum_{k=0}^{n} p_{\mathrm{sgn}(Z)}^{(n-k)}(X-\mathrm{sgn}(Z)|-\mathrm{sgn}(Z))\mathcal{M}^{(k)}_{\nu} (Z)
\label{eq:singlevacpropa}
\end{align}
with $\mathcal{M}_{\nu}^{(n)}(Z)$ the probability that the vacancy starting in $Z$ exchanged position with the TP in time $n$ knowing that it started with polarity $\nu$. From Equation (\ref{eq:singlevacpropa}), we see that the single vacancy propagator only depends on $\mathcal{M}^{(n)}_{\nu}(Z)$ and the propagator in the case of a vacancy starting right next to the RTP. Denoting $F_{\mathrm{sgn}(\mu\nu)} ^{(n)}$ the first-passage time (FPT) density in $X=0$ for a vacancy starting in $\mu = \pm1$, knowing that the RTP started with a polarity $\nu$ at $n=0$, we notice that

\begin{equation}
\mathcal{M}^{(n)}_{\nu}(Z) = \sum_{k=0}^{n} f_{\mathrm{sgn}(Z)Z-1}^{(k)} \left[ S^{(k)}_{+}F_{\mathrm{sgn}(\nu Z)}^{(n-k)}+ S^{(k)}_{-}F_{-\mathrm{sgn}(\nu Z)}^{(n-k)}\right]
\end{equation}
where $f_X^{(n)}$ denotes the classical first-passage time density at the origin at time $n$ of a symmetrical one-dimensional P\'olya walk starting at $n=0$ at position $X$ \citep{hughes-book-1995} and $S^{(n)}_{\mathrm{sgn}(\mu \nu)}$ the probability for the RTP to have a given polarity $\mu$ at timestep $n$, knowing that it was $\nu$ at $n=0$ for which it is straightforward to obtain exact expressions (see details in \citep{Note1}).
In the particular case of a vacancy starting in $\mu = \pm 1$, this probability reads (see details in \cite{Note1})
\begin{align}
&p_{\nu}^{(n)} (X|\mu) =  \delta_{X,0} \left( 1-\sum_{j=1}^n F_{\mathrm{sgn(\mu\nu)}} ^{(j)} \right) + \sum_{j=1}^{\infty} \sum_{m_1,\cdots,m_j}^{\infty}  \nonumber \\ 
&\times  \delta_{\sum_{i=1}^{j} m_k,n} \delta_{X,\mu[(-1)^j+1]/2}  F_{\mathrm{sgn(\mu\nu)}} ^{(m_1)} F_{-} ^{(m_2)} \cdots F_{-} ^{(m_{j-1})} \nonumber \\
&\times \left( 1-\sum_{i=1}^{m_{j}} F_{-} ^{(i)} \right)
\label{eq:singlevacprob}
\end{align}
with $\delta_{i,j}$ the Kronecker delta. The first term in the right-hand side of  Equation (\ref{eq:singlevacprob}) gives the probability that at time $n$ the RTP was never visited by the vacancy. The second term represents a partition over the number of visits by the vacancy $j$ and the waiting times between successive visits $m_j$. Here, the key step of our derivation is therefore to compute the first-passage time densities $F_{\mathrm{sgn}(\mu\nu)} ^{(n)}$ which a priori couple the polarity dynamics and vacancy dynamics. Thanks to the Markovian dynamics of the RTP polarity, we obtain exact full expressions for these FPT densities and their generating functions as shown in \citep{Note1}. For instance, we show that $F_{-} ^{(n)}$  obeys the following recurrence relation,
\begin{align}
&F^{(n)}_{-} = \alpha \left\{ (1-q_1)\delta_{n,1} + q_{1} \sum_{k=1}^n \left[ f_1^{(k-1)} S^{(k-1)}_{+} F^{(n-k)}_{+} \right. \right. \nonumber \\
		&\left. \left. + f_1^{(k-1)} S^{(k-1)}_{-} F^{(n-k)}_{-}\right] \right\} + \alpha \sum_{k=1}^n \left[ f_1^{(k-1)} S^{(k-1)}_{-} F^{(n-k)}_{+} \right.\nonumber \\
		&  \left.+ f_1^{(k-1)} S^{(k-1)}_{-} F^{(n-k)}_{-}\right].
\label{eq:FTP1}
\end{align}

Defining the generating function of any time-dependent function $g^{(n)}$ as $\widehat{g}(\xi) \equiv \sum_{n=0}^{\infty} g^{(n)}\xi^n$, Equations (\ref{eq:singlevacpropa})-(\ref{eq:singlevacprob}) implie that the generating function of the single-vacancy propagator can be written in terms of the generating functions of the first-passage time densities above as
\begin{equation}
\widehat{p}_{\nu}(X|\mu; \xi) = \frac{\delta_{X,0}\left[1-\delta_{\mathrm{sgn}(\mu),\nu} (\widehat{F}_{+}-\widehat{F}_{-})\right] + \delta_{X,\mu}\widehat{F}_{-}}{(1-\xi)(1+\widehat{F}_{-})}
\end{equation}
where we have used the short-hand notations $\widehat{F}_{\pm} = \widehat{F}_{\pm}(\xi)$ and the original position of the vacancy, $\mu = \pm1$.

{\it Systems with a small concentration of vacancies ---} 
We now consider a lattice containing a low concentration of vacancies $\rho_0$; we assume that the lattice of size $2L$ contains $M$ vacancies such that $\rho_0 \equiv M/2L$. First, we consider the case of a fixed initial polarity for the RTP, $\nu$. Following \citep{brummelhuis-physa-1989}, we write $P_{\nu}^{(n)}(X| \{Z_j\}) $ the probability that the RTP with original polarity $\nu$ is at position $X$ at time $n$ provided that the $M$ vacancies started at positions $\{Z_j\}_{j \in [1,M]}$. At the lowest order in the vacancies density, we consider the contributions of each vacancy to be independent and we write
\begin{align}
P_{\nu}^{(n)}(X| \{Z_j&\}) = \sum_{Y_1,...,Y_M} \delta_{X,Y_1+\cdots+Y_M} P_{\nu}^{(n)}(\{Y_j\} | \{Z_j\})  \nonumber \\
				   &\underrel{\rho_0 \to 0}{=} \sum_{Y_1,...,Y_M} \delta_{X,Y_1+\cdots+Y_M} \prod_{j=1}^M p_{\nu}^{(n)}(Y_j | Z_j)
\end{align}
where $P^{(n)}(\{ Y_j\}|\{ Z_j\})$ is the conditional probability that in time $n$ the RTP has performed a displacement $Y_1$ due to interaction with the vacancy 1, $Y_2$ due to interaction with the vacancy 2 etc. We define the Fourier transform in space as $\mathcal{F}\{X\} \equiv X^*(q) = \sum_{y} \mathrm{e}^{iqy}X(y)$, where the sum runs over all lattice sites. In Fourier space and averaged over the initial distribution of vacancies positions (assumed to be uniformly distributed on the lattice, except for the origin which is occupied by the RTP and denoted $\bar{\cdot}$~), we obtain
\begin{equation}
\bar{P}^*_{\nu}(q,n) = \left [\bar{p}^*_{\nu}(q,n) \right]^M \underrel{\rho_0 \to 0}{\sim} 1 - \rho_0\Omega_{\nu}(q,n)
\end{equation}
where we impose $L,M \to \infty$ (while $\rho_0$ is kept constant) and we define
\begin{align}
\Omega_{\nu}(q,n)& = \sum_{k=0}^{n} \left[ [1-p^*_{+}(q|-1;n-k)\mathrm{e}^{iq}]\sum_{Z=1}^{+\infty}\mathcal{M}^{(k)}_{\nu}(Z) \right. \nonumber \\
				& \left. +[1-p^*_{-}(q|+1;n-k)\mathrm{e}^{-iq}] \sum_{Z=-\infty}^{-1}\mathcal{M}^{(k)}_{\nu}(Z) \right].
\end{align}

Lastly, we average over the initial polarity of the RTP to obtain
\begin{equation}
\bar{P}^*(q,n) = \frac{1}{2}\left [\bar{P}^*_{+}(q,n) + \bar{P}^*_{-}(q,n)  \right] \underrel{\rho_0 \to 0}{\sim} 1 - \rho_0\Omega(q,n)
\end{equation}
with $\Omega(q,n) \equiv [\Omega_+(q,n)+\Omega_-(q,n)]/2$. The generating function of the second characteristic function, which is defined as $ \psi(q,\xi) \equiv \sum_{n=0}^{\infty} \ln\left< \mathrm{e}^{iqX_n}\right>\xi^n$,
satisfies the following relation
\begin{equation}
\lim_{\rho_0 \to 0} \frac{\psi(q,\xi)}{\rho_0} = -\widehat{\Omega}(q,\xi)
\end{equation}
After lengthy but straightforward algebra (see details in \cite{Note1}), we proceed to an expansion of $\widehat{\Omega}(q,\xi)$ in power series of $q$ to obtain
\begin{equation}
\widehat{\Omega}(q,\xi) = -\frac{\sum_{n=1}^\infty \left[1+(-1)^n \right]\left[ \widehat{F}_+ +\widehat{F}_-\right](iq)^n/n! }{2(1-\xi)(1-(1-\sqrt{1-\xi^2})/\xi)(1+\widehat{F}_-)}.
\end{equation}

We now have all the tools to derive our central analytical result which defines the exact (in the leading order in $\rho_0$) cumulant generating function. The cumulants $\kappa_j$ of arbitrary order $j$ are defined by $\ln\left[ \bar{P}^*(q,n) \right] \equiv \sum_{j=1}^{\infty} \kappa_j^{(n)} (iq)^j /j! $. We can identify equal order terms in both expressions and find
\begin{equation}
\widehat{\kappa}_j(\xi) \underrel{\rho_0 \to 0}{=}  \frac{\rho_0}{2} \frac{\left[1+(-1)^j \right]\left[ \widehat{F}_+(\xi) +\widehat{F}_-(\xi)\right]}{(1-\xi)(1-(1-\sqrt{1-\xi^2})/\xi)(1+\widehat{F}_-(\xi))} 
\label{eq:cumulants}
\end{equation}

Recalling that the FPT densities $\widehat{F}_{\nu}(\xi)$ are explicitly given in \cite{Note1} in terms of the tumbling probability (appearing explicitly through the Poissonian dynamics of the polarity), Equation (\ref{eq:cumulants}) provides an expression for all cumulants exact in the leading order in $\rho_0$  in Laplace space. Strikingly, all cumulants of same parity are equal and in particular, all odd cumulants are identically equal to 0. While we easily understand why odd cumulants in a process averaged over the initial polarity should be zero, it is interesting to note that in the case where we fix the initial polarity of the RTP, the odd cumulants were non-zero but decaying as $1/\sqrt{n}$ (see details in \citep{Note1}). As a side note, we show in \cite{Note1} that our predictions retrieve the results for the infinitely biased tracer particle in the $\alpha \to 0$ limit \citep{illien-prl-2013}.

{\it Cumulants in the long-time limit ---} We expand the generating function of the even cumulants in power series of $1-\xi$ (which is equivalent to a long-time limit expansion) 
\begin{equation}
\lim_{\rho_0 \to 0} \frac{\widehat{\kappa}_{\mathrm{even}}(\xi)}{\rho_0} \underrel{\xi \to 1}{=} \frac{1}{\sqrt{2}} \frac{1}{(1-\xi)^{3/2}} - \frac{\sqrt{\alpha}}{2\sqrt{1-\alpha}} \frac{1}{1-\xi}+\mathcal{O}\left(1\right).
\end{equation}

\begin{figure}[t]
\centering
\includegraphics[width=0.5\textwidth]{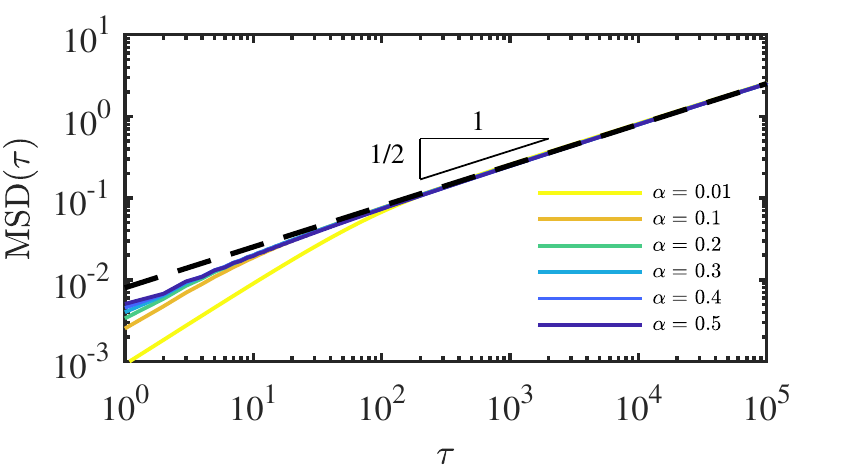}
\caption{{\it Mean square displacement of the RTP} - shown for various tumbling probabilities $\alpha \in [0.01;0.5]$ showing an $\alpha$-dependent transient to a subdiffusive long-time scaling independent of $\alpha$; at long times, $\mathrm{MSD}(n) \to \sqrt{2n/\pi}$ (dashed black line).}
\label{fig:msds}
\end{figure}

Using Tauberian theorems \citep{feller-book-1971}, we can invert term by term this expression and we obtain in the time domain
\begin{equation}
\lim_{\rho_0 \to 0} \frac{\kappa_{\mathrm{even}}^{(n)}}{\rho_0} \underrel{n\to\infty}{=} \sqrt{\frac{2n}{\pi}} -  \frac{\sqrt{\alpha}}{2\sqrt{1-\alpha}}  + o(n^{-1/2}).
\label{eq:cumulantexpansion}
\end{equation}

\begin{figure}[t]
\centering
\includegraphics[width=0.45\textwidth]{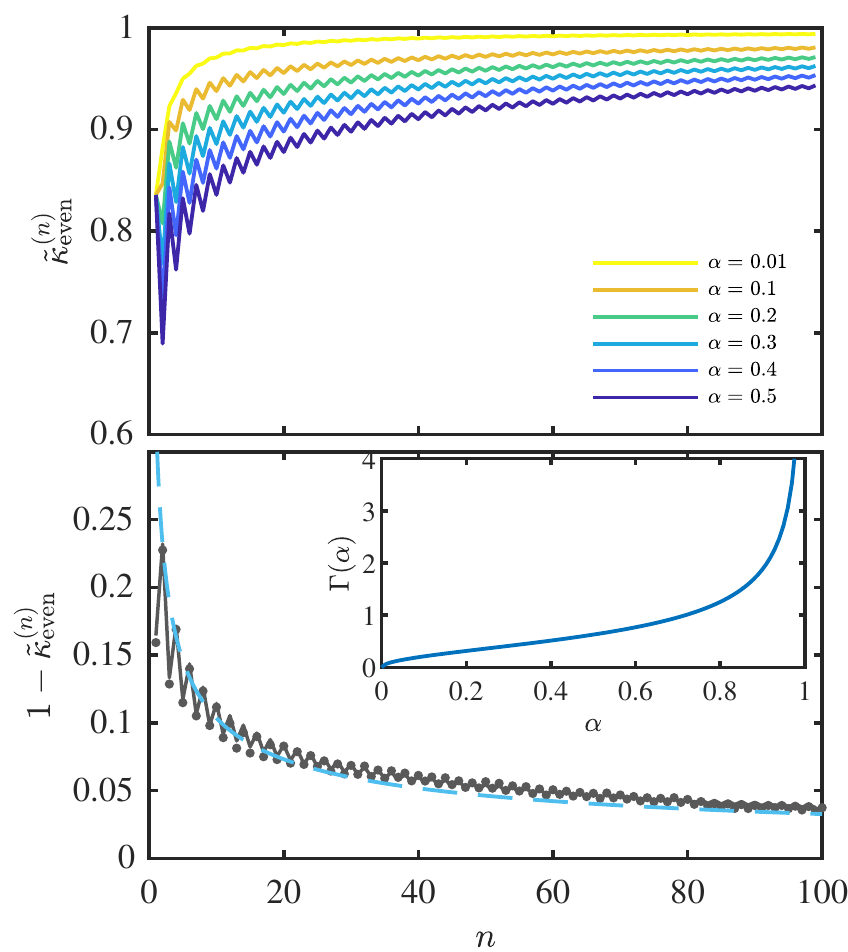}
\caption{{\it Reduced cumulants} - ({\it Top}) Reduced cumulants $\tilde{\kappa}_{\mathrm{even}}^{(n)} = \kappa_{\mathrm{even}}^{(n)}/(\rho_0\sqrt{2n/\pi})$ vs time $n$ for several tumbling probabilities, showing transients dependent on $\alpha$ (increasing $\alpha$ from yellow to blue) obtained from the inversion of Equation (\ref{eq:cumulants}); ({\it Bottom}) Reduced cumulant for a tumbling probability $\alpha = 0.3$ and a vacancies density of $\rho_0 = 0.001$, from inversion of Equation (\ref{eq:cumulants}) (solid line), numerical simulations (symbols) and first higher order term in the Taylor series of the cumulants as in Equation (\ref{eq:cumulantexpansion}) (dashed line), $1-\tilde{\kappa}_{\mathrm{even}}^{(n)} = \Gamma(\alpha)/\sqrt{n}$. The $\alpha$-dependance of the higher-order term is shown in inset.}
\label{fig:cumulants}
\end{figure}

From Equation (\ref{eq:cumulantexpansion}), we first notice that the leading asymptotics are surprisingly independent of the tumbling probability $\alpha$. In Figure \ref{fig:msds}, we show that the mean-square displacement of the RTP converges in the long-time limit to the anomalous subdiffusive scaling observed for passive single-file diffusion. This comes from the fact that, for all finite values of $\alpha$, the statistics of the waiting times between successive steps of the RTP is asymptotically independent of $\alpha$. Secondly, by obtaining higher order terms in the expansion of the cumulants, we realize that $\alpha$ enters in the first subleading order term with the expected monotonicity. As shown in Figure \ref{fig:cumulants}, as the tumbling probability decreases ({\it i.e.} the more persistent the RTP becomes), the importance of the higher order terms in the expansion decreases and the dynamics of the RTP converges faster to the classical single-file scaling.

Surprisingly, we observe that the time-averaged MSD decreases faster for smaller $\alpha$ for short timescales $t\ll1/\alpha$ (see Figure \ref{fig:msds}). This counterintuitive result is opposite to the monotonicity expected in the absence of obstacles. After this transient, one recovers the expected monotonicity of the MSD with respect to tumbling probability for $t\gg1/\alpha$. Thus, the time-averaged stationary mean-square displacement (Figure \ref{fig:msds}) shows an inversion of monotonicity of the second cumulant compared to that of the ensemble-averaged cumulants (Figure \ref{fig:cumulants}) in the $n \to 0$ limit, a common signature of aging phenomena. Namely, this result suggests that the stationary distribution of vacancy positions is different from the Poissonian initial conditions we consider in the derivation leading to Figure \ref{fig:cumulants}. Indeed, the bath particles density displays fluctuations in the vicinity of the RTP (with an increase of particles density in front of the RTP and a decrease at the back). This implies that a higher persistence can lead to a slowing down of the dynamics at short timescales, a counterintuitive idea reminiscent of the concept of negative differential mobility observed for biased tracers and run-and-tumble particles in crowded environments \citep{benichou-prl-2014,bertrand-prl-2018,baerts-pre-2013}.

\begin{figure}[t]
\centering
\includegraphics[width=0.5\textwidth]{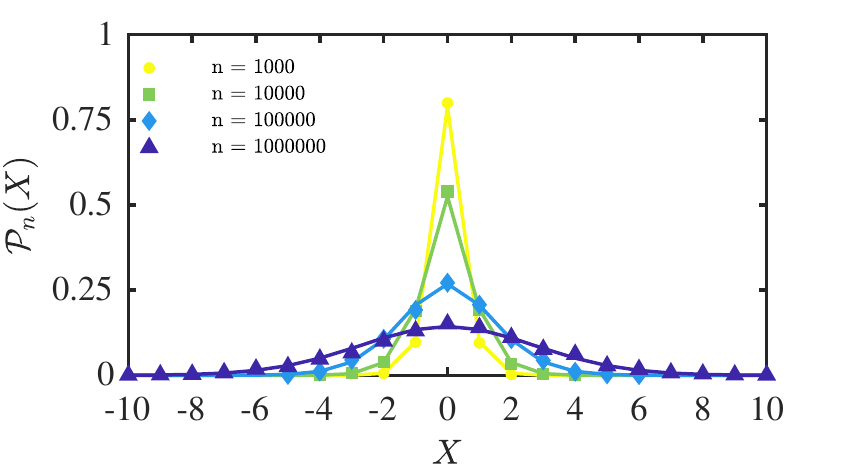}
\caption{{\it Distribution of positions of a RTP} with random initial polarity $\mathcal{P}_n(X)$, tumbling probability $\alpha = 0.2$ and vacancy density $\rho_0 = 0.01$ for various times $n=10^3$, $10^4$, $10^5$ and $10^6$ (from yellow to blue), for numerical simulations (symbols) and theoretical predictions obtained from Equation \ref{eq:distributionsBiasedRT} (solid lines).}
\label{fig:distributions}
\end{figure}

{\it Full statistics of the RTP position ---} Finally, the equality of same parity cumulants (to leading order in $\rho_0$) implies that the associated distribution is of Skellam type \citep{skellam-jrss-1946}. As a consequence, our results provide the complete distribution of RTP positions $\mathcal{P}_n(X)$ for any time $n$, which in our case simplifies to
\begin{equation}
\mathcal{P}_n(X) \underrel{\rho_0 \to 0}{\sim} \exp\left( -\kappa_{\mathrm{even}}^{(n)} \right) I_X \left( \left| \kappa_{\mathrm{even}}^{(n)}\right| \right),
\label{eq:distributionsBiasedRT}
\end{equation}
where $I_n$ is a modified Bessel function of the first kind \citep{abramowitz-book-1972}. Figure \ref{fig:distributions} shows a perfect agreement between the predicted distributions and our numerical simulations. In the long time limits, we recover the results derived by \citet{levitt-pra-1973} and \citet{beijeren-prb-1983}. Importantly, we find that independently of the tumbling probability the rescaled variable $X/(2\rho_0^2n/\pi)^{1/4}$ is asymptotically distributed according to a normal law.

{\it Summary ---} Using a lattice model, we derived expressions for the full statistics of positions of run-and-tumble particles with arbitrary tumbling probability $\alpha$ in dense single-file environment. Our predictions are exact to the leading order in vacancy density $\rho_0 \to 0$. We have shown that the asymptotic dynamics of the RTP displays an anomalous subdiffusive scaling $\propto \sqrt{n}$ with prefactor independent of $\alpha$. Further, we highlighted the presence of aging in this system for which the stationary distribution of bath particles is not Poissonian but displays density fluctuations in the vicinity of the RTP leading strikingly to a slowing down of the short timescale dynamics for more persistent RTPs. Run-and-tumble particles are a canonical model of natural and artificial self-propelled particles; as a consequence, we believe that these exact results will find a wealth of applications; they set the ground for a quantitative analysis of experimental trajectories of real biological or artificial microswimmers in extreme confinement. 


%

\onecolumngrid
\pagebreak
\begin{center}
\textbf{\large Supplemental Material for "Dynamics of a Run-and-Tumble Particle in Dense Diffusive Single-File Systems"}
\end{center}
\setcounter{equation}{0}
\setcounter{figure}{0}
\setcounter{table}{0}
\setcounter{page}{1}
\makeatletter
\renewcommand{\theequation}{S\arabic{equation}}
\renewcommand{\thefigure}{S\arabic{figure}}
\renewcommand{\bibnumfmt}[1]{[S#1]}
\renewcommand{\citenumfont}[1]{#1}

\section{Hopping probabilities}
\label{sec:hopping}

While previous studies examined the dynamics of {\it externally driven} tracers \citep{illien-prl-2013,benichou-prl-2014}, we interested here in a model of {\it internally driven} self-propelled particles. We consider a one-dimensional lattice infinite in both directions. The lattice sites are occupied by hard-core particles present at mean density $\rho$ performing symmetrical random walks, with the restriction that the occupancy number for each site is at most equal to one. The mean density of vacancies is thus equal to $\rho_0 = 1-\rho$. The tracer particle (TP) performs a biased random walk and is initially placed at the origin. The jump direction is chosen with probability $1/2$ for the gas particles in the bulk. While the TP chooses to hop along the biasing direction with probability $p_+$, and in the opposite direction with probability $p_{-}$, such that $p_{+}+p_{-}=1$. At each time step, the TP reverses its {\it polarity} (direction of the bias) with probability $\alpha$. This model is an example of {\it Run-and-Tumble motion} \cite{berg-book-2004}.

We consider that the bias originates from the application of a {\it constant internal drive} $\mathbf{F} = F \mathbf{e}$ applied to the run-and-tumble particle (RTP) in the direction of its polarity. In this case, we know that the probability for the RTP to be in position $x$ in steady state is given by a Boltzmann distribution: 
\begin{equation}
P_{\mathrm{stat}} (x) = \frac{1}{\mathcal{Z}} \mathrm{e}^{\beta F x},
\end{equation}
where $\mathcal{Z}$ is a normalization constant and $\beta = 1/(k_BT)$ is the inverse temperature. Let us consider for instance that the polarity of the RTP points towards the positive integers ($\mathbf{e} = \hat{x}$). If we denote, $p(1 \rightarrow 0)$ the probability for the RTP to go from $x=1$ to $x=0$ in one step, we know that the detailed balance condition imposes that: 
\begin{equation}
P_{\mathrm{stat}}(0) p( 0 \rightarrow 1) = P_{\mathrm{stat}}(1) p( 1 \rightarrow 0)
\end{equation}

We know that by definition: $p( 0 \rightarrow 1) = p_{+}$ and $p( 1 \rightarrow 0) = p_{-}$. Thus, one gets that: 
\begin{equation}
\frac{p_+}{p_{-}} = \mathrm{e}^{\beta F}
\end{equation}

In order to fulfill the detailed balance condition, we finally have that the hopping probabilities are given by:
\begin{equation}
p_{+} = \frac{\mathrm{e}^{\frac{1}{2} \beta F}}{\mathrm{e}^{\frac{1}{2} \beta F}+\mathrm{e}^{-\frac{1}{2} \beta F}},~p_{-} = \frac{\mathrm{e}^{-\frac{1}{2} \beta F}}{\mathrm{e}^{\frac{1}{2} \beta F}+\mathrm{e}^{-\frac{1}{2} \beta F}}
\end{equation}

In the high density limit, the TP only moves when a vacancy visits one of its neighboring sites. Hence, it is interesting to look at the dynamics of the vacancies rather than the dynamics of the particles. In our one-dimensional problem, a vacancy moves at each step according to the following evolution rules: 
\begin{itemize}[topsep=0em,noitemsep]
\item if the vacancy occupies the site to the right of the TP, it has a probability $q_1 = 1/(2p_1+1)$ to jump to the right and $1-q_1$ to jump to the left;
\item if the vacancy occupies the site to the left of the TP, it has a probability $q_{-1} = 1/(2p_{-1}+1)$ to jump to the left and $1-q_{-1}$ to jump to the right;
\item in any other case, the vacancy has a probability 1/2 to jump to the left and 1/2 to jump to the right.
\end{itemize}

with by definition $p_1 = p_+$ and $p_{-1} = p_-$ (respectively, $p_1 = p_+$ and $p_{-1} = p_-$) if the polarity points in the positive direction (respectively, in the negative direction). According to those definitions, the vacancies perform a P\'olya walk everywhere on the lattice except in the vicinity of the RTP \citep{hughes-book-1995}. 

In this Letter, we consider only the case of an infinitely biased RTP ($F\to \infty$) along a polarity direction which flips on exponentially distributed times. In the case the polarity of the RTP points in the positive direction, the evolution rules thus read:
\begin{itemize}[topsep=0em,noitemsep]
\item if the vacancy occupies the site to the right of the RTP, it has a probability $q_1 = 1/3$ to jump to the right and $1-q_1=2/3$ to jump to the left (exchange its position with the RTP);
\item if the vacancy occupies the site to the left of the RTP, it has a probability $q_{-1} = 1$ to jump to the left and $1-q_{-1}=0$ to jump to the right (exchange its position with the RTP);
\item in any other case, the vacancy has a probability 1/2 to jump to the left and 1/2 to jump to the right.
\end{itemize}
We can easily obtain the evolutions rules in the case the polarity points in the reverse direction by symmetry.

\section{Polarity dynamics of the Run-and-Tumble particle}
\label{sec:polarity}

The polarity dynamics of the RTP is decoupled from its motion and is solely controlled by its flipping probability $\alpha$. We know that the polarity of our RTP can be positive ($+$) or negative ($-$). We will denote  $S^{(k)}_{\mu \nu}$ the probability for the TP to have a given polarity $\mu$ at time step $k$, knowing that it was $\nu$ at $k=0$. For all time $k$ assuming that we started with a positive polarity, we can write the following master equation: 
\begin{align}
S^{(k)}_{++} &= \alpha S^{(k-1)}_{-+} + (1-\alpha)S^{(k-1)}_{++} \\
S^{(k)}_{-+} &= 1 - S^{(k)}_{++}
\end{align}

Using the fixed-points method to solve this reccurence relation, we obtain that for all times $k$,
\begin{align}
S^{(k)}_{++} &= \frac{1}{2}\left[1+(1-2\alpha)^k \right]\\
S^{(k)}_{-+} &= 1-S^{(k)}_{++}
\end{align}

We can similarly obtain the following probabilities in the case we start with a negative polarity. In all generality, we can write for a polarity $\mu = \{\pm\}$ that: 
\begin{equation}
S^{(k)}_{\mu \mu} = \frac{1}{2}\left[1+(1-2\alpha)^k \right] = 1-S^{(k)}_{-\mu \mu}.
\end{equation}

For the sake of simplicity, we will sometimes write $S^{(k)}_{\mu \mu} = S^{(k)}_{+}$ and $S^{(k)}_{-\mu \mu} = S^{(k)}_{-}$ in what follows.

\section{Single vacancy propagator}
\label{sec:singlevac}

Following \citet{brummelhuis-jstatphys-1988}, we first consider an auxiliary problem: a system containing a single vacancy. We consider a lattice on the integers $l\in[-L;L]$, the vacancy can occupy any site but the origin. As we saw in the main text, we can write the probability to find the RTP in position $X$ at time $n$, knowing it started with polarity $\nu$ at time $n=0$ as an average over the initial condition for the position of the vacancy $Z$,
\begin{equation}
\bar{p}_\nu^{(n)}(X) = \frac{1}{2L} \left(\sum_{Z=-L}^{-1} p_\nu^{(n)}(X|Z) + \sum_{Z=1}^{L} p_\nu^{(n)}(X|Z)\right)
\end{equation}

with $p_\nu^{(n)}(X|Z)$ defined (as in the main text) as the probability to find the RTP in position $X$ at a time $n$ knowing its polarity was initially $\nu$ and that the vacancy started in $Z$. In this particular case, the RTP can only be found in two positions depending on the location of the vacancy at $n=0$: 
\begin{itemize}[topsep=0em,noitemsep]
\item if the vacancy starts on the right of the RTP, the RTP can be in 0 or 1;
\item if the vacancy starts on the left of the RTP, the RTP can be in -1 or 0.
\end{itemize}

First, we consider the case where $Z>0$ and the RTP started with a positive polarity at $n=0$. We can write the following recurrence relation: 
\begin{equation}
p_+^{(n)}(X|Z) = \delta_{X,0} \left( 1 - \sum_{k=0}^n \mathcal{M}^{(k)}_+(Z)\right) + \sum_{k=0}^{n}   p_+^{(n-k)}(X-1|-1) \mathcal{M}^{(k)}_+(Z)
\label{Seq:propfromZ1}
\end{equation}
with $\mathcal{M}^{(n)}_{+}(Z)$ the probability that the vacancy starting in $Z$ exchanged positions with the TP in time $n$ knowing that its polarity was $+$ at $n=0$. In particular, we notice that $\mathcal{M}^{(n)}_{+}(Z)$ can be written as the convolution of the first-passage time density of a P\'olya walk to the adjacent site to the RTP and the first-passage time density of exchange of RTP and vacancy position, for a vacancy starting on one of the lattice sites adjacent to the RTP knowing the polarity of the RTP.
\begin{equation}
\mathcal{M}^{(n)}_{+}(Z) = 
\left\lbrace
\begin{array}{ll}
\sum_{k=0}^n  f_{Z-1}^{(k)} S_{++}^{(k)} \mathcal{F}_{++}^{(n-k)} + \sum_{k=0}^n  f_{Z-1}^{(k)} S_{-+}^{(k)} \mathcal{F}_{+-}^{(n-k)}& \mathrm{for~} Z>0\\
\sum_{k=0}^n  f_{-1-Z}^{(k)} S_{++}^{(k)} \mathcal{F}_{-+}^{(n-k)} + \sum_{k=0}^n  f_{Z-1}^{(k)} S_{-+}^{(k)} \mathcal{F}_{--}^{(n-k)} & \mathrm{for~} Z<0
\end{array}\right.
\end{equation}
where $f^{(n)}_X$ denotes the first-passage time density at the origin at time $n$ of a symmetrical one-dimensional P\'olya walk starting in position $X$ at $n=0$ and $\mathcal{F}^{(n)}_{\mu\nu}$ is the first-passage time density of exchange of the positions of the RTP and the vacancy in time $n$, knowing that the RTP started with an initial polarity $\nu$ and the vacancy started in $Z=\mu1$. Full expressions for those FTP densities are provided in Section \ref{sec:FTP}. We note that we have similarly
\begin{equation}
\mathcal{M}^{(n)}_{-}(Z) = 
\left\lbrace
\begin{array}{ll}
\sum_{k=0}^n  f_{Z-1}^{(k)} S_{+-}^{(k)} \mathcal{F}_{++}^{(n-k)} + \sum_{k=0}^n  f_{Z-1}^{(k)} S_{--}^{(k)} \mathcal{F}_{+-}^{(n-k)}& \mathrm{for~} Z>0\\
\sum_{k=0}^n  f_{-1-Z}^{(k)} S_{+-}^{(k)} \mathcal{F}_{-+}^{(n-k)} + \sum_{k=0}^n  f_{Z-1}^{(k)} S_{--}^{(k)} \mathcal{F}_{--}^{(n-k)} & \mathrm{for~} Z<0
\end{array}\right.
\end{equation}

In Equation (\ref{Seq:propfromZ1}), the first term corresponds to the probability that the RTP has not moved and the second term is composed of a convolution between the probability to have exchanged positions in $k$ steps and the probability that the TP travels $X-1$ steps in time $n-k$, granted that it had a polarity $+$ at $n=0$ and that the vacancy started now in $-1$ (vacancy and RTP exchanged positions).

Similarly in the case $Z<0$, we have: 
\begin{equation}
p_+^{(n)}(X|Z) = \delta_{X,0} \left( 1 - \sum_{k=0}^n \mathcal{M}^{(k)}_+(Z)\right) + \sum_{k=0}^{n}  p_-^{(n-k)}(X+1|+1) \mathcal{M}^{(k)}_+(Z)
\label{Seq:propfromZ2}
\end{equation} 

(in this expression the second term contains the probability $p_-^{(n-k)}(X+1|+1)$ because we know that the propagator needs to be shifted following interaction with the vacancy coming from the left of the RTP and the fact that the polarity had to flip in order for the RTP and the vacancy to exchange positions.)

By symmetry, we obtain similarly, 
\begin{equation}
p_-^{(n)}(X|Z) = 
\begin{cases}  
\delta_{X,0} \left( 1 - \sum_{k=0}^n \mathcal{M}^{(k)}_-(Z)\right) + \sum_{k=0}^{n}   p_+^{(n-k)}(X-1|-1) \mathcal{M}^{(k)}_-(Z), & \text{if } Z>0 \\
\delta_{X,0} \left( 1 - \sum_{k=0}^n \mathcal{M}^{(k)}_-(Z)\right) + \sum_{k=0}^{n}  p_-^{(n-k)}(X+1|+1)  \mathcal{M}^{(k)}_-(Z), & \text{if } Z<0
\end{cases}
\label{Seq:propfromZ3}
\end{equation}

As we can see from Equations (\ref{Seq:propfromZ1}), (\ref{Seq:propfromZ2}) and (\ref{Seq:propfromZ3}), finding the general single vacancy propagator reduces to finding an expression for the single vacancy propagator in the case of a vacancy adjacent to the RTP. With some care, the general single vacancy propagator can be expressed using the first-passage statistics the vacancy starting on a site adjacent to the RTP to its position. For instance, the probability to find the RTP in position $X$ in time $n$ knowing that the TP started in $X=0$ with a positive polarity and a vacancy in $Z=+1$ reads: 
\begin{equation}
\begin{aligned}
p_+^{(n)}(X|+1) = & \delta_{X,0} \left(1-\sum_{j=0}^n \mathcal{F}_{++}^{(j)} \right)  \\
		  + & \delta_{X,0} \sum_{j=1}^{\infty} \sum_{m_1=0}^{\infty} \ldots \sum_{m_{2j+1}=0}^{\infty} \delta_{\sum_{i=1}^{2j+1}m_i,n}\mathcal{F}_{++}^{(m_1)}\mathcal{F}_{-+}^{(m_2)}\mathcal{F}_{+-}^{(m_3)} \ldots \mathcal{F}_{-+}^{(m_{2j})} \left(1-\sum_{k=0}^{m_{2j+1}} \mathcal{F}_{+-}^{(k)} \right) \\
		 +  & \delta_{X,1} \sum_{j=1}^{\infty} \sum_{m_1=0}^{\infty} \ldots \sum_{m_{2j}=0}^{\infty} \delta_{\sum_{i=1}^{2j}m_i,n} \mathcal{F}_{++}^{(m_1)}\mathcal{F}_{-+}^{(m_2)}\mathcal{F}_{+-}^{(m_3)}\ldots \mathcal{F}_{+-}^{(m_{2j-1})}\left(1-\sum_{k=0}^{m_{2j}} \mathcal{F}_{-+}^{(k)} \right)
\end{aligned}
\label{Seq:svpropagatorRT+}
\end{equation}

Equation (\ref{Seq:svpropagatorRT+}) is composed of three terms:
\begin{enumerate}[topsep=0em,noitemsep]
\item The RTP is in 0, it was never visited by the vacancy;
\item The RTP is in 0, it was visited an {\it even} number of times; this term contains $2j$ visits and $1$ last non-visiting event;
\item The RTP is in 1, it was visited an {\it odd} number of times; this term contains $2j+1$ visits and $1$ last non-visiting event.
\end{enumerate}

Taking the discrete Laplace transform of Equation (\ref{Seq:svpropagatorRT+}) and denoting the FPT densities $\widehat{\mathcal{F}}_{\mu \nu}(\xi) =\widehat{F}_{\mathrm{sgn}(\mu \nu)}$ for the sake of simplicity, we find
\begin{equation}
\widehat{p}_+(X|+1;\xi) =  \frac{1}{1-\xi} \left[ \delta_{X,0} (1-\widehat{F}_+) + \delta_{X,0} (1-\widehat{F}_-) \sum_{j=1}^{\infty} \widehat{F}_+\widehat{F}_-^{2j-1} + \delta_{X,1} (1-\widehat{F}_-) \sum_{j=1}^{\infty} \widehat{F}_+\widehat{F}_-^{2j-2} \right]
\end{equation}
which finally reduces to, 
\begin{equation}
\widehat{p}_+(X|+1;\xi)  =  \frac{\delta_{X,0} (1-\widehat{F}_+ + \widehat{F}_-)+\delta_{X,1}\widehat{F}_+}{(1-\xi)(1+\widehat{F}_-)}.
\end{equation}

Similarly, we find that
\begin{equation}
\begin{aligned}
p_+^{(n)}(X|-1) = & \delta_{X,0} \left(1-\sum_{j=0}^n \mathcal{F}_{-+}^{(j)} \right) + \\
		  & \delta_{X,0} \sum_{j=1}^{\infty} \sum_{m_1=0}^{\infty} \ldots \sum_{m_{2j+1}=0}^{\infty} \delta_{\sum_{i=1}^{2j+1}m_i,n} \mathcal{F}_{-+}^{(m_1)}\mathcal{F}_{+-}^{(m_2)}\mathcal{F}_{-+}^{(m_3)}\ldots \mathcal{F}_{+-}^{(m_{2j})}\left(1-\sum_{k=0}^{m_{2j+1}} \mathcal{F}_{-+}^{(k)} \right) + \\
		  & \delta_{X,-1} \sum_{j=1}^{\infty} \sum_{m_1=0}^{\infty} \ldots \sum_{m_{2j}=0}^{\infty} \delta_{\sum_{i=1}^{2j}m_i,n} \mathcal{F}_{-+}^{(m_1)}\mathcal{F}_{+-}^{(m_2)}\mathcal{F}_{-+}^{(m_3)}\ldots \mathcal{F}_{-+}^{(m_{2j-1})}\left(1-\sum_{k=0}^{m_{2j}} \mathcal{F}_{+-}^{(k)} \right)
\end{aligned}
\label{Seq:svpropagatorRT-}
\end{equation}

Taking the discrete Laplace transform of Equation (\ref{Seq:svpropagatorRT-}), we obtain:
\begin{equation}
\widehat{p}_+(X|-1;\xi)  =  \frac{\delta_{X,0} +\delta_{X,-1}\widehat{F}_-}{(1-\xi)(1+\widehat{F}_-)}
\label{Seq:sprop1}
\end{equation}

By symmetry, we can write the remaining two cases,
\begin{align}
\widehat{p}_-(X|+1;\xi)  &=  \frac{\delta_{X,0} +\delta_{X,1}\widehat{F}_-}{(1-\xi)(1+\widehat{F}_-)} \label{Seq:sprop2}\\
\widehat{p}_-(X|-1;\xi)  &=   \frac{\delta_{X,0} (1-\widehat{F}_+ + \widehat{F}_-)+\delta_{X,-1}\widehat{F}_+}{(1-\xi)(1+\widehat{F}_-)}
\end{align}

Finally, we realize that the single vacancy propagator only depends on the FPT densities $\mathcal{F}^{(n)}_{\mu \nu}$, 
\begin{equation}
\begin{aligned}
\bar{p}_\nu^{(n)}(X) = & \delta_{X,0} \left( 1 - \frac{1}{2L}\sum_{Z\ne0}\sum_{k=0}^n \mathcal{M}^{(k)}_\nu(Z)\right)  \\
			      &+ \frac{1}{2L}\sum_{k=0}^{n} p_-^{(n-k)}(X+1|+1)\sum_{Z=-L}^{-1}\mathcal{M}^{(k)}_\nu(Z) +\frac{1}{2L}\sum_{k=0}^{n} p_+^{(n-k)}(X-1|-1)\sum_{Z=1}^{L}\mathcal{M}^{(k)}_\nu(Z).
\end{aligned}
\label{Seq:singlevacpropa}
\end{equation}
We provide a full derivation of those quantities in Section \ref{sec:FTP}.

\section{First-passage time densities for vacancies adjacent to the RTP}
\label{sec:FTP}

Knowing that the RTP polarity is in a given state at $n=0$ and that a vacancy is adjacent to the TP at that time, we want to calculate the probability that a first interaction ({\it i.e.} exchange of positions) will happen at time $n$. In all generality, there are four possible configurations to consider at $n=0$ assuming that the RTP is in $X=0$: 
\begin{itemize}[topsep=0em,noitemsep]
\item RTP polarity is  ($+$) and the vacancy is in $X=+1$;
\item RTP polarity is  ($+$) and the vacancy is in $X=-1$;
\item RTP polarity is  ($-$) and the vacancy is in $X=+1$;
\item RTP polarity is  ($-$) and the vacancy is in $X=-1$. 
\end{itemize}

We denote $\mathcal{F}_{\mu \nu}^{(n)}$ the first-passage time density in the case where the RTP polarity is originally $\nu$ and the vacancy started in $Z=\mu1$. We will detail here the derivation on an example. We will consider that the polarity is positive at $n=0$. Knowing that a vacancy is next to the RTP at $n=0$, we want to calculate the probability of first interaction (exchange of positions) at time $n$. In this particular case, the vacancy has a chance to interact with the RTP at the first time step, and we can express this quantity via the First-Passage Time (FPT) density at the origin at time $n$ of a symmetrical one-dimensional P\'{o}lya walk starting at $n=0$ at position $l$ and denoted $f_l^{(n)}$. Thus, we write the FPT density as the following convolution: 
\begin{align}
\mathcal{F}^{(n)}_{++} = (1-\alpha) & \left\{ (1-q_1)\delta_{n,1} + q_{1} \sum_{k=1}^n \left[ f_1^{(k-1)} S^{(k-1)}_{++} \mathcal{F}^{(n-k)}_{++} + f_1^{(k-1)} S^{(k-1)}_{-+} \mathcal{F}^{(n-k)}_{+-}\right] \right\}\nonumber \\
				     & + \alpha \sum_{k=1}^n \left[ f_1^{(k-1)} S^{(k-1)}_{+-} \mathcal{F}^{(n-k)}_{++} + f_1^{(k-1)} S^{(k-1)}_{--} \mathcal{F}^{(n-k)}_{+-}\right].
\label{Seq:FTP1}
\end{align}
At each time step, the RTP (i) flips its polarity with probability $\alpha$ and (ii) attempts to make a step in the direction of its polarity. The first term in equation (\ref{Seq:FTP1}) corresponds to the case where the RTP does not flip its polarity at $n=1$. In this case, we know that the RTP has a chance to exchange positions with the vacancy at $n=1$ (this is the first term in the curly brackets), the second term is based on the probability that the vacancy did not exchange positions with the RTP at $n=1$ but came back in $k-1$ steps while the polarity is still positive, and the third term is based on the probability that the vacancy did not exchange positions with the RTP at $n=1$ but came back in $k-1$ steps while the polarity flipped in the meantime. The second term in equation (\ref{Seq:FTP1}) corresponds to the case where the RTP does flip its polarity at $n=1$. This case is similar to the first term, with the exception that we know that the RTP and the adjacent vacancy cannot interact at the first time step. 

Similarly, for a vacancy starting on the left of the RTP with a positive polarity, we write 
\begin{align}
\mathcal{F}^{(n)}_{-+} = \alpha & \left\{ (1-q_1)\delta_{n,1} + q_{1} \sum_{k=1}^n \left[ f_1^{(k-1)} S^{(k-1)}_{--} \mathcal{F}^{(n-k)}_{--} + f_1^{(k-1)} S^{(k-1)}_{+-} \mathcal{F}^{(n-k)}_{-+}\right] \right\}\nonumber \\
				     & +(1- \alpha) \sum_{k=1}^n \left[ f_1^{(k-1)} S^{(k-1)}_{-+} \mathcal{F}^{(n-k)}_{--} + f_1^{(k-1)} S^{(k-1)}_{++} \mathcal{F}^{(n-k)}_{-+}\right].
\label{Seq:FTP2}
\end{align}

For a given vacancy position, the polarity of the TP can be pointing towards the vacancy or not. By symmetry, we only need to compute these two cases. We denote $F^{(n)}_+$ the First-Time Passage density for the case where the vacancy is on the right side of the RTP given its polarity, while we denote $F^{(n)}_-$ the inverse case and we have,
\begin{align}
\mathcal{F}^{(n)}_{++} = \mathcal{F}^{(n)}_{--} = F^{(n)}_+ \\
\mathcal{F}^{(n)}_{-+} = \mathcal{F}^{(n)}_{+-} = F^{(n)}_-
\end{align}

Furthermore, by definition, we can also write that $f_l^{(n)} = f_{-l}^{(n)}$. For the sake of simplicity, we will define :
\begin{align}
g_+^{(n)} = f^{(n-1)}_1S^{(n-1)}_{++}= f^{(n-1)}_1S^{(n-1)}_{--}\\
g_-^{(n)} = f^{(n-1)}_1S^{(n-1)}_{-+} = f^{(n-1)}_1S^{(n-1)}_{+-}
\end{align}

We can study the discrete Laplace transform of these two quantities, defined as: $\widehat{F}(\xi) = \sum_{n=0}^{\infty} \xi^n F(n)$. Using the convolution theorem, we obtain: 
\begin{align}
\widehat{F}_{+}(\xi) &= (1-\alpha)\left\{ (1-q_1)\xi + q_1 \left[ \widehat{g}_{+}(\xi) \widehat{F}_{+}(\xi) + \widehat{g}_{-}(\xi) \widehat{F}_{-}(\xi)\right] \right\} + \alpha \left[ \widehat{g}_{-}(\xi) \widehat{F}_{+}(\xi)+ \widehat{g}_{+}(\xi) \widehat{F}_{-}(\xi) \right] \label{Seq:laplaceFTP1}\\
\widehat{F}_{-}(\xi) &= \alpha\left\{ (1-q_1)\xi + q_1 \left[ \widehat{g}_{+}(\xi) \widehat{F}_{+}(\xi) + \widehat{g}_{-}(\xi) \widehat{F}_{-}(\xi)\right] \right\} + (1- \alpha) \left[ \widehat{g}_{-}(\xi) \widehat{F}_{+}(\xi)+ \widehat{g}_{+}(\xi) \widehat{F}_{-}(\xi) \right] \label{Seq:laplaceFTP2}
\end{align}

We can combine equations (\ref{Seq:laplaceFTP1}) and (\ref{Seq:laplaceFTP2}) to obtain finally
\begin{align}
\widehat{F}_{+}(\xi) &= \frac{(1-q_1)\left[1-\alpha - (1-2\alpha)\widehat{g}_{+}(\xi))\right] \xi}{1-(1+q_1)\left[ (1-\alpha)\widehat{g}_{+}(\xi)+\alpha \widehat{g}_{-}(\xi)\right] + q_1(1-2\alpha)\left[\widehat{g}_{+}^2(\xi) - \widehat{g}_{-}^2(\xi) \right]} \label{Seq:FTPRT1} \\
\widehat{F}_{-}(\xi) &= \frac{(1-q_1)\left[\alpha + (1-2\alpha)\widehat{g}_{-}(\xi))\right] \xi}{1-(1+q_1)\left[ (1-\alpha)\widehat{g}_{+}(\xi)+\alpha \widehat{g}_{-}(\xi)\right] + q_1(1-2\alpha)\left[\widehat{g}_{+}^2(\xi) - \widehat{g}_{-}^2(\xi) \right]} \label{Seq:FTPRT2} \\
\end{align}

Going back to the definition of the discrete Laplace transform, we obtain the following expressions
\begin{align}
\widehat{g}_{+}(\xi) &= \sum_{n=1}^{\infty} \xi^n f_1^{(n-1)}(1+(1-2\alpha)^{n-1})/2 = \frac{\xi}{2}\left[ \widehat{f}_1(\xi)+\widehat{f}_1((1-2\alpha)\xi)\right] \\
\widehat{g}_{-}(\xi) &= \sum_{n=1}^{\infty} \xi^n f_1^{(n-1)}(1-(1-2\alpha)^{n-1})/2  = \frac{\xi}{2}\left[ \widehat{f}_1(\xi)-\widehat{f}_1((1-2\alpha)\xi)\right] \\
\end{align}

with by definition of a P\'{o}lya walk: 
\begin{equation}
\widehat{f}_x(\xi) = \left( \frac{1-\sqrt{1-\xi^2}}{\xi} \right)^{|x|}
\end{equation}

As a conclusion, we have obtained exact expressions for the First-Passage Time densities we needed to complete our expression of the single vacancy propagator.

\section{Single File with a small concentration of vacancies}
\label{sec:conditioning}

In this section, we consider now the case of a small but finite concentration of vacacancies, we assume that the system contains $M$ vacancies such that $\rho_0 = M/2L$. We will start by deriving expression for the cumulants, exact in the linear order in the density of vacancies. We will then show that our results are consistent with the results in the case of a biased tracer particle (derived in Reference \cite{illien-prl-2013}). Finally, we will generalize our derivation to the case of a random initial polarity to obtain the cumulants and full statistics for the RTP position.

\subsection{Case of a fixed initial polarity}

First, we consider the case where we fix the initial polarity of the RTP. Following Brummelhuis and Hilhorst \citep{brummelhuis-jstatphys-1988,brummelhuis-physa-1989}, we write in general $P^{(n)}_{\nu}(X|\{Z_j\})$ the probability that the RTP is at position $X$ at time $n$ provided that the $M$ vacancies were at positions $\{ Z_j\}_{j \in [1,M]}$ and the RTP polarity was $\nu$ at $n=0$. We can write: 
\begin{equation}
P^{(n)}_{\nu}(X|\{Z_j\}) = \sum_{Y_1,\ldots, Y_M} \delta_{X,Y_1+\ldots Y_M} P^{(n)}_{\nu}(\{ Y_j\}|\{ Z_j\})
\end{equation}
where $P^{(n)}_{\nu}(\{ Y_j\}|\{ Z_j\})$ is the conditional probability that, within the time interval $n$, the RTP has performed a displacement $Y_1$ due to interaction with the vacancy 1, $Y_2$ due to interaction with the vacancy 2 etc. In the lowest order of the vacancy density $\rho_0$, the vacancies contribute independently to the displacement:
\begin{equation}
P^{(n)}_{\nu}(\{ Y_j\}|\{ Z_j\}) \underrel{\rho_0 \rightarrow 0}{=} \prod_{j=1}^M p^{(n)}_{\nu}(Y_j|Z_j)
\end{equation}

Thus, we can express this probability as a function of the single vacancy propagator:
\begin{equation}
P^{(n)}_{\nu}(X|\{ Z_j\}) \underrel{\rho_0 \rightarrow 0}{=} \sum_{Y_1,\ldots, Y_M} \delta_{X,Y_1+\ldots Y_M} \prod_{j=1}^M p^{(n)}_{\nu}(Y_j|Z_j)
\end{equation}

If we suppose that the vacancies are uniformly distributed, we can average $P^{(n)}_{\nu}(X|\{ Z_j\})$ over the initial distribution of vacancies:
\begin{align}
\bar{P}^{(n)}_{\nu}(X) & = \overline{\sum_{Y_1,\ldots, Y_M} \delta_{X,Y_1+\ldots Y_M} \prod_{j=1}^M p^{(n)}_{\nu}(Y_j|Z_j) } \\
				   & = \sum_{Y_1,\ldots, Y_M} \delta_{X,Y_1+\ldots Y_M} \overline{\prod_{j=1}^M p^{(n)}_{\nu}(Y_j|Z_j) } \\
				   &  \underrel{\rho_0 \rightarrow 0}{=}  \sum_{Y_1,\ldots, Y_M} \delta_{X,Y_1+\ldots Y_M} \prod_{j=1}^M \bar{p}^{(n)}_{\nu}(Y_j)
\label{Seq:averagespinconditioning}
\end{align}

By definition, the Fourier transform in space is written as: 
\begin{equation}
\mathcal{F}\{ X \} = X^*(q) = \sum_{y=-\infty}^{+\infty} \mathrm{e}^{iqy} X(y)
\end{equation}
where the sum runs over all lattice sites.

For the sake of simplicity and without loss of generality, we can assume that the spin is always positive in $n=0$. As a consequence of Equation (\ref{Seq:averagespinconditioning}), the probability averaged over initial conditions for the vacancies reduces in Fourier transform to
\begin{equation}
\bar{P}^*_{+}(q,n) = \left [\bar{p}^*_{+}(q,n) \right]^M,
\label{Seq:avpos}
\end{equation}
where we see that the total contribution of the $M$ vacancies reduces to a superposition of the contributions of single vacancies. This expression gives formally a relationship between the general propagator and the single vacancy propagator. From Equation (\ref{Seq:singlevacpropa}), we write in Fourier space: 
\begin{equation}
\bar{p}_+^{*}(q,n) = 1 - \frac{1}{2L} \sum_{k=0}^{n} \left[ [1-p^*_-(q|+1;n-k)\mathrm{e}^{-iq}] \sum_{Z=-L}^{-1}\mathcal{M}^{(k)}_+(Z) +[1-p^*_+(q|-1;n-k)\mathrm{e}^{iq}]\sum_{Z=1}^{L}\mathcal{M}^{(k)}_+(Z) \right]
\end{equation}

We can rewrite Equation (\ref{Seq:avpos}) as:

\begin{equation}
\bar{P}^*_{+}(q;n) = \left[ 1 - \frac{1}{2L}\Omega^L_+(q,n) \right]^M,
\label{Seq:barP2}
\end{equation}

where we define the following quantity
\begin{equation}
\Omega^L_{+}(q,n;L) = \sum_{k=0}^{n} \left[ [1-p^*_-(q|+1;n-k)\mathrm{e}^{-iq}] \sum_{Z=-L}^{-1}\mathcal{M}^{(k)}_+(Z) +[1-p^*_+(q|-1;n-k)\mathrm{e}^{iq}]\sum_{Z=1}^{+L}\mathcal{M}^{(k)}_+(Z) \right] .
\end{equation}

By definition of the Fourier transform for a random variable $X_n$ : 
\begin{equation}
\bar{P}^*_{+}(q,n) = \left< \mathrm{e}^{iqX_n }\right>
\end{equation}

The cumulant generating function, defined as $\psi_n(q) = \ln \left< \mathrm{e}^{iqX_n } \right>$, reads in the limit of low vacancies density ($\rho_0 \ll 1$):
\begin{equation}
\psi_n(q)  \equiv \ln \bar{P}^*_{+}(q,n) \underrel{\rho_0 \to 0}{\sim} -\rho_0 \Omega^{\infty}_+(q,n)
\end{equation}

with $\rho_0 = M/2L$ and $M,L \to \infty$. This leads to the Z-transform relation: 

\begin{equation}
\lim_{\rho_0 \to 0} \frac{\psi (q,\xi)}{\rho_0} = -\sum_{k=0}^{\infty} \Omega^{\infty}_+(q,n)\xi^k = -\widehat{\Omega}_+(q,\xi)
\end{equation}

By discrete Laplace transform, we obtain: 

\begin{equation}
\widehat{\Omega}_+(q,\xi) = \left[ \frac{1}{1-\xi} -\widehat{p}^*_-(q|+1,\xi)\mathrm{e}^{-iq}\right] h_{-}(\xi) + \left[ \frac{1}{1-\xi} -\widehat{p}^*_+(q|-1,\xi)\mathrm{e}^{iq}\right]h_{+}(\xi)
\label{Seq:fourieromega}
\end{equation}

with $h_{\mu} = \sum_{Z=\mu1}^{\mu\infty} \widehat{\mathcal{M}}_+(\xi)$, we provide full calculation and expressions for these quantitites in Section \ref{sec:hmu}.

\subsection{Calculation of $h_{\mu}$ }
\label{sec:hmu}

We have already noticed that: 
\begin{equation}
\mathcal{M}^{(n)}_{+}(Z) = 
\left\lbrace
\begin{array}{ll}
\sum_{k=0}^n  f_{Z-1}^{(k)} S_{++}^{(k)} \mathcal{F}_{++}^{(n-k)} + \sum_{k=0}^n  f_{Z-1}^{(k)} S_{-+}^{(k)} \mathcal{F}_{+-}^{(n-k)}& \mathrm{for~} Z>0\\
\sum_{k=0}^n  f_{-1-Z}^{(k)} S_{++}^{(k)} \mathcal{F}_{-+}^{(n-k)} + \sum_{k=0}^n  f_{-1-Z}^{(k)} S_{-+}^{(k)} \mathcal{F}_{--}^{(n-k)} & \mathrm{for~} Z<0
\end{array}\right.
\end{equation}

The associated generating functions are thus simply: 
\begin{equation}
\widehat{\mathcal{M}}_{+}(Z,\xi) = 
\left\lbrace
\begin{array}{ll}
\widehat{f_{Z-1} S}_{++}(\xi) \widehat{F}_{+}(\xi) + \widehat{f_{Z-1} S}_{-+}(\xi) \widehat{F}_{-}(\xi) & \mathrm{for~} Z>0\\
\widehat{f_{-1-Z} S}_{++}(\xi) \widehat{F}_{-}(\xi) + \widehat{f_{-1-Z} S}_{-+}(\xi) \widehat{F}_{+}(\xi) & \mathrm{for~} Z<0
\end{array}\right.
\end{equation}

In particular, we have
\begin{align}
\widehat{f_XS_{++}}(\xi) &= \frac{1}{2}\left[ \widehat{f}_X(\xi) + \widehat{f}_X(\xi(1-2\alpha))\right] \\
\widehat{f_XS_{-+}}(\xi) &= \frac{1}{2}\left[ \widehat{f}_X(\xi) - \widehat{f}_X(\xi(1-2\alpha))\right]
\end{align}

Finally, we can write that: 
\begin{align}
h_{+}(\xi) &= \sum_{Z=1}^{\infty} \widehat{\mathcal{M}}_{+}(Z,\xi) \nonumber \\
	       &= \frac{1}{2}\left[ \frac{\widehat{F}_+(\xi) + \widehat{F}_-(\xi)}{1-\widehat{f}_1(\xi)} + \frac{\widehat{F}_+(\xi) - \widehat{F}_-(\xi)}{1-\widehat{f}_1(\xi(1-2\alpha))}\right]
\end{align}

We can also write: 
\begin{align}
h_{-}(\xi) &= \sum_{Z=-\infty}^{-1} \widehat{\mathcal{M}}_{+}(Z,\xi) \nonumber \\
	      &= \frac{1}{2}\left[ \frac{\widehat{F}_+(\xi) + \widehat{F}_-(\xi)}{1-\widehat{f}_1(\xi)} - \frac{\widehat{F}_+(\xi) - \widehat{F}_-(\xi)}{1-\widehat{f}_1(\xi(1-2\alpha))}\right]
\end{align}

In conclusion, in general, we have: 
\begin{equation}
h_{\mu}(\xi) = \frac{1}{2}\left[ \frac{\widehat{F}_+(\xi) + \widehat{F}_-(\xi)}{1-\widehat{f}_1(\xi)} +\mathrm{sgn}(\mu) \frac{\widehat{F}_+(\xi) - \widehat{F}_-(\xi)}{1-\widehat{f}_1(\xi(1-2\alpha))}\right]
\label{Seq:hmu}
\end{equation}

\subsection{Expression for the cumulants with positive initial polarity}

From Equations (\ref{Seq:sprop1}) and (\ref{Seq:sprop2}), the Laplace-Fourier transform of the single vacancy propagator is given by: 
\begin{align}
\widehat{p}^*_+(q|-1;\xi)  &=  \frac{1 +\mathrm{e}^{-iq}\widehat{F}_-(\xi)}{(1-\xi)(1+\widehat{F}_-(\xi))} \label{Seq:fouriersvprop1}\\
\widehat{p}^*_-(q|+1;\xi)  &=  \frac{1 +\mathrm{e}^{iq}\widehat{F}_-(\xi)}{(1-\xi)(1+\widehat{F}_-(\xi))}\label{Seq:fouriersvprop2}
\end{align}

Combining Equations (\ref{Seq:fourieromega}), (\ref{Seq:hmu}), (\ref{Seq:fouriersvprop1}) and (\ref{Seq:fouriersvprop2}),   we finally obtain: 
\begin{align}
\widehat{\Omega}_+(q,\xi) = &\frac{1-\mathrm{e}^{-iq}}{2[1-\xi][1+\widehat{F}_-(\xi)]} \left[ \frac{\widehat{F}_+(\xi) + \widehat{F}_-(\xi)}{1-\widehat{f}_1(\xi)} - \frac{\widehat{F}_+(\xi) - \widehat{F}_-(\xi)}{1-\widehat{f}_1(\xi(1-2\alpha))} \right] + \nonumber \\
					   &\frac{1-\mathrm{e}^{iq}}{2[1-\xi][1+\widehat{F}_-(\xi)]} \left[ \frac{\widehat{F}_+(\xi) + \widehat{F}_-(\xi)}{1-\widehat{f}_1(\xi)} + \frac{\widehat{F}_+(\xi) - \widehat{F}_-(\xi)}{1-\widehat{f}_1(\xi(1-2\alpha))} \right]
\end{align}

On one hand, we can proceed to the expansion of $\widehat{\Omega}_+(q,\xi)$ in power series of $q$: 
\begin{align}
\widehat{\Omega}_+(q,\xi) = &-\frac{\widehat{F}_+(\xi) + \widehat{F}_-(\xi)}{2[1-\xi][1+\widehat{F}_-(\xi)][1-\widehat{f}_1(\xi)]}  \sum_{j=1}^{\infty} \frac{(iq)^j}{j!} \left[ 1+(-1)^j \right]  \nonumber \\ 
					    &-\frac{\widehat{F}_+(\xi) - \widehat{F}_-(\xi)}{2[1-\xi][1+\widehat{F}_-(\xi)][1-\widehat{f}_1(\xi(1-2\alpha))]}  \sum_{j=1}^{\infty} \frac{(iq)^j}{j!} \left[ 1-(-1)^j \right]
\end{align}

Recalling the definition of the generating functions of the cumulants $\kappa^{(n)}_{j}$of arbitrary order $j$, we can write
\begin{equation}
\psi_n(q) = \ln \bar{P}^*_{+}(q,n) \equiv \sum_{j=1}^{\infty} \frac{\kappa_j^{(n)}}{j!} (iq)^j = -\rho_0 \Omega_+(q,n)
\end{equation}

So we can identify same order terms and write that: 
\begin{equation}
\widehat{\kappa}_j(\xi) \underrel{\rho_0 \to 0}{=}  \frac{\rho_0}{2[1-\xi][1+\widehat{F}_-(\xi)]} \left\{ \frac{\widehat{F}_+(\xi) + \widehat{F}_-(\xi)}{1-\widehat{f}_1(\xi)} \left[ 1+(-1)^j \right] + \frac{\widehat{F}_+(\xi) - \widehat{F}_-(\xi)}{1-\widehat{f}_1(\xi(1-2\alpha))} \left[ 1-(-1)^j \right] \right\}
\label{Seq:cumulantsRTpos}
\end{equation}

Equation (\ref{Seq:cumulantsRTpos}) provides an exact expression of the cumulants in the Fourier-Laplace space. Recalling that the functions $\widehat{F}_{\nu}(\xi)$ are explicitly given in Section \ref{sec:FTP} in terms of the tumbling probability, this equation gives an expression of the cumulants of arbitrary order.

\subsection{Cumulants in the long-time limit}

From Equation (\ref{Seq:cumulantsRTpos}), we see that all odd cumulants have the same generating function $\widehat{\kappa}_{\mathrm{odd}}(\xi)$ and all even cumulants have the same generating function $\widehat{\kappa}_{\mathrm{even}}(\xi)$. We recall that the expression for the $\widehat{F}_{\nu}(\xi)$ is given by Equations  (\ref{Seq:FTPRT1}) and (\ref{Seq:FTPRT2}). We can thus proceed to an expansion in power series of $1-\xi$ (which is equivalent to a long-time expansion in the time domain) of the generating function of the cumulants and using the fact that $q_1 = 1/3$, we find that: 

\begin{align}
& \lim_{\rho_0 \to 0} \frac{\widehat{\kappa}_{\mathrm{odd}}(\xi)}{\rho_0} \underrel{\xi \to 1}{=}  \frac{\sqrt{2}(1-2\alpha)^2}{4\sqrt{\alpha(1-\alpha)} \left[ \sqrt{\alpha(1-\alpha)}-\alpha \right] \sqrt{1-\xi}}+ \mathcal{O}\left( 1\right) \\
& \lim_{\rho_0 \to 0} \frac{\widehat{\kappa}_{\mathrm{even}}(\xi)}{\rho_0} \underrel{\xi \to 1}{=} \frac{1}{\sqrt{2}} \frac{1}{(1-\xi)^{3/2}} +\mathcal{O}\left( \frac{1}{\sqrt{1-\xi}}\right)
\end{align}

At this point, it is useful to remember the Tauberian theorem \citep{feller-book-1971}. For a time-dependent function $\phi(n)$ and its associated generating function $\widehat{\phi}(\xi) = \sum_{n=1}^{\infty} \phi(n) \xi^n$, if the expansion of $\widehat{\phi}(\xi)$ in powers of $(1-\xi)$ has the form 
\begin{equation}
\widehat{\phi}(\xi) \underrel{\xi \to 1}{\sim}\frac{1}{(1-\xi)^\chi}\Phi\left(\frac{1}{1-\xi} \right),
\end{equation}

Then, the long time behavior of $\phi(n)$ is given by
\begin{equation}
\phi(t) \underrel{n \to \infty}{\sim} \frac{1}{\Gamma(\chi)}n^{\chi-1}\Phi(n),
\end{equation}

where $\Gamma$ is the usual gamma function. This relation holds if $\chi>0$, $\phi(n)>0$, $\phi(n)$ is monotonic and $\Phi$ is slowly varying in the sense that 
\begin{equation}
\lim_{x\to\infty} \frac{\Phi(\lambda x)}{\Phi(x)} = 1
\end{equation}
for any $\lambda >0$.

Using the Tauberian theorem, we find that the long-time behavior of the odd cumulants is given by 

\begin{equation}
\lim_{\rho_0 \to 0} \frac{\kappa_{\mathrm{odd}}(n)}{\rho_0} \underrel{n\to\infty}{=} \frac{(1-2\alpha)^2}{\sqrt{8\pi n}\sqrt{\alpha(1-\alpha)} \left[ \sqrt{\alpha(1-\alpha)}-\alpha \right]}+o(1/n).
\end{equation}

and a long-time behavior of the even cumulants given by
\begin{equation}
\lim_{\rho_0 \to 0} \frac{\kappa_{\mathrm{even}}(n)}{\rho_0} \underrel{n\to\infty}{=} \sqrt{\frac{2n}{\pi}}+o(1).
\end{equation}

We note that remarkably the leading order in time of the even cumulants does not depend on $\alpha$, while the leading order in time of the odd cumulants does and decays to zero in the long time limit. The asymptotic behavior of $\kappa_{\mathrm{even}}(n)$ tells us that the variance of the RTP position grows as $\sqrt{n}$, this subdiffusive behavior is the one obtained for the classical symmetric single file dynamics.

\subsection{Retrieving the case of a biased TP in the $\alpha \to 0$ limit}

For sanity, we can check our calculation against the result for a biased TP \citep{illien-prl-2013}. In this section, we will check that our derivation for a run-and-tumble tracer particle in the limit $\alpha \to 0$ gives the same prediction as in the case of an infinitely biased tracer particle. In particular, we recall that the cumulants of all order in the biased case are given by:
\begin{equation}
\lim_{\rho_0 \to 0 } \frac{\widehat{\kappa}^b_j(\xi)}{\rho_0} = \frac{\widehat{F}_1(1-\widehat{F}_{-1}) + (-1)^j\widehat{F}_{-1}(1-\widehat{F}_1)}{(1-\xi)(1-(1-\sqrt{1-\xi^2})/\xi)(1-\widehat{F}_1\widehat{F}_{-1})}
\end{equation}
where $\widehat{F}_{\pm1} = (1-q_{\pm1}\xi)/(1-q_{\pm1}(1-\sqrt{1-\xi^2}))$ and $q_{\pm1}$ defined as in the RTP case. In the case of an infinite bias, we know that $q_1 = 1/3$ and $q_{-1}=1$ leading to
\begin{equation}
\lim_{\rho_0 \to 0 } \frac{\widehat{\kappa}^b_j(\xi)}{\rho_0} = \frac{2\xi}{(1-\xi)(1-(1-\sqrt{1-\xi^2})/\xi)(2+\sqrt{1-\xi^2})}
\end{equation}

In the Run-and-Tumble case, we need to first take the limit of low tumbling rate, $\alpha \to 0$, before taking the limit of long-times. Using Equations (\ref{Seq:FTPRT1}) and (\ref{Seq:FTPRT2}) with $q_1=1/3$, we can write that the FTP densities as:
\begin{align}
& \widehat{F}_+ (\xi) = \frac{2\xi}{2+\sqrt{1-\xi^2}} \\
& \widehat{F}_- (\xi) = 0
\end{align}

Injecting this result in Equation (\ref{Seq:cumulantsRTpos}), we obtain that for all orders: 
\begin{equation}
\lim_{\rho_0 \to 0 } \frac{\widehat{\kappa}_j(\xi)}{\rho_0} = \frac{2\xi}{(1-\xi)(1-(1-\sqrt{1-\xi^2})/\xi)(2+\sqrt{1-\xi^2})}
\end{equation}

We notice that: (1) cumulants in the infinitely biased case and the RTP case in the limit $\alpha \to 0$ are equal and (2) expressions for the cumulants for all orders are equal. This means that in all cases, all cumulants are equal and in particular, an expansion in power series of $1-\xi$ gives:
\begin{align}
& \lim_{\rho_0 \to 0} \frac{\widehat{\kappa}_j(\xi)}{\rho_0} \underrel{\xi \to 1}{=} \frac{1}{\sqrt{2}} \frac{1}{(1-\xi)^{3/2}} + \mathcal{O}\left( \frac{1}{\sqrt{1-\xi}}\right) \\
\end{align}

Hence, we conveniently retrieve the biased case in the zero tumbling rate limit ($\alpha \to 0$).

\section{General case: random initial spin}
\label{sec:random}

We now generalize the results of the previous section to the more general case, where the polarity is not fixed to a particular direction at $n=0$. For that, we need to average over trajectories conditioned with positive and negative initial polarity. 

\subsection{Cumulants in the case of a negative initial polarity}
\label{sec:negativecond}

It is rather straightforward to calculate the specific case of a fixed negative polarity and check that it gives us the expected result considering the derivation in section \ref{sec:conditioning}. Similarly to the previous derivation, we write in this case,

\begin{equation}
\bar{P}^*_{-}(q,n) = \left [\bar{p}^*_{-}(q,n) \right]^M
\label{Seq:avneg}
\end{equation}

From Equation (\ref{Seq:singlevacpropa}), we can write:
\begin{equation}
\bar{p}_-^{*}(q,n) = 1 - \frac{1}{2L} \sum_{k=0}^{n} \left[ [1-p^*_-(q|+1;n-k)\mathrm{e}^{iq}] \sum_{Z=-L}^{-1}\mathcal{M}^{(k)}_-(Z) +[1-p^*_+(q|-1;n-k)\mathrm{e}^{-iq}]\sum_{Z=1}^{L}\mathcal{M}^{(k)}_-(Z) \right]
\end{equation}

Following the same procedure and definition as in Section \ref{sec:conditioning}, the previous expression expanded reads
\begin{align}
\widehat{\Omega}_-(q,\xi) = &-\frac{\widehat{F}_+(\xi) + \widehat{F}_-(\xi)}{2[1-\xi][1+\widehat{F}_-(\xi)][1-\widehat{f}_1(\xi)]}  \sum_{j=1}^{\infty} \frac{(iq)^j}{j!} \left[ 1+(-1)^j \right]  \nonumber \\ 
					    &+\frac{\widehat{F}_+(\xi) - \widehat{F}_-(\xi)}{2[1-\xi][1+\widehat{F}_-(\xi)][1-\widehat{f}_1(\xi(1-2\alpha))]}  \sum_{j=1}^{\infty} \frac{(iq)^j}{j!} \left[ 1-(-1)^j \right]
\end{align}

Finally the cumulants are given by identification of the $n$-th order terms in the development, and we obtain:
\begin{equation}
\widehat{\kappa}_j(\xi) \underrel{\rho_0 \to 0}{=}  \frac{\rho_0}{2[1-\xi][1+\widehat{F}_-(\xi)]} \left\{ \frac{\widehat{F}_+(\xi) + \widehat{F}_-(\xi)}{1-\widehat{f}_1(\xi)} \left[ 1+(-1)^j \right] - \frac{\widehat{F}_+(\xi) - \widehat{F}_-(\xi)}{1-\widehat{f}_1(\xi(1-2\alpha))} \left[ 1-(-1)^j \right] \right\}
\label{Seq:cumulantsRTneg}
\end{equation}

Comparing Equations (\ref{Seq:cumulantsRTpos}) and (\ref{Seq:cumulantsRTneg}), it is easy to see that these expressions will yield the same cumulants for all even orders, and cumulants with opposite signs for all odd orders.

\subsection{Expression of the cumulants in the general case}

The final step is now to average over the initial spin, we condition on the spin right after the averaging over initial conditions for a small concentration of vacancies. As such, we write: 

\begin{equation}
\bar{P}^*(q,n) = \frac{1}{2} \left[ \bar{P}^*_{+}(q,n) + \bar{P}^*_{-}(q,n)\right] 
\end{equation}

Reinjecting in this expression equations (\ref{Seq:avpos}) and (\ref{Seq:avneg}), we obtain
\begin{equation}
\bar{P}^*(q,n)  = \frac{1}{2} \left[ \left[\bar{p}^*_{+}(q,n) \right]^M + \left [\bar{p}^*_{-}(q,n) \right]^M \right] \underrel{\rho_0 \to 0}{\sim} 1 - \rho_0\Omega^L(q,n)		       		       
\end{equation}

with $\Omega^{\infty}(q,n) = \left[\Omega^{\infty}_+(q,n) +\Omega^{\infty}_-(q,n) \right]/2$ when $L,M \to \infty$. In the Laplace domain, we obtain then

\begin{equation}
\widehat{\Omega}(q,\xi) = \frac{1}{2}\left[\widehat{\Omega}_+(q,\xi) +\widehat{\Omega}_-(q,\xi) \right]
\end{equation}

{\it i.e.} 
\begin{equation}
\widehat{\Omega}(q,\xi) = -\frac{\widehat{F}_+(\xi) + \widehat{F}_-(\xi)}{2[1-\xi][1+\widehat{F}_-(\xi)][1-\widehat{f}_1(\xi)]}  \sum_{j=1}^{\infty} \frac{(iq)^j}{j!} \left[ 1+(-1)^j \right]
\end{equation}

As usual, the expression for the cumulants is given by identification of the terms in the expansion and we get
\begin{equation}
\widehat{\kappa}_j(\xi) \underrel{\rho_0 \to 0}{=}  \frac{ \rho_0}{2}\frac{ \widehat{F}_+(\xi) + \widehat{F}_-(\xi)   }{(1-\xi)(1-\widehat{f}_1(\xi))(1+\widehat{F}_-(\xi)) } \left[ 1+(-1)^j \right]
\label{Seq:cumulantsRT}
\end{equation}

As a conclusion, we can easily see that all even cumulants are equal, as are all odd cumulants. We obtain the following final cumulants
\begin{align}
& \lim_{\rho_0 \to 0} \frac{\widehat{\kappa}_{\mathrm{even}}(\xi)}{\rho_0} =  \frac{\widehat{F}_+(\xi) + \widehat{F}_-(\xi)}{(1-\xi)(1-(1-\sqrt{1-\xi^2})/\xi)(1+\widehat{F}_-(\xi))} \\
& \lim_{\rho_0 \to 0} \frac{\widehat{\kappa}_{\mathrm{odd}}(\xi)}{\rho_0} = 0 
\end{align}

We realize that all odd cumulants are identically equal to 0. As for the even cumulants, they present the same form as the even cumulants in the case of a positive conditioning, which was expected as the initial polarity only affects the dynamics until the first polarity flip. We can thus proceed to an expansion in power series of $1-\xi$ the generating function of the cumulant and using the fact that $q_1 = 1/3$, we can proceed to a Taylor expansion of the cumulants at long time to higher orders and we obtain:
\begin{equation}
\lim_{\rho_0 \to 0} \frac{\widehat{\kappa}_{\mathrm{even}}(\xi)}{\rho_0} \underrel{\xi \to 1}{=} \frac{1}{\sqrt{2}} \frac{1}{(1-\xi)^{3/2}} - \frac{\sqrt{\alpha}}{2\sqrt{1-\alpha}} \frac{1}{1-\xi} + \mathcal{O}( 1)
\end{equation}

This expression can be inverted term by term and we obtain in time domain 
\begin{equation}
\lim_{\rho_0 \to 0} \frac{\kappa_{\mathrm{even}}(n)}{\rho_0} \underrel{n\to\infty}{=} \sqrt{\frac{2n}{\pi}} -\frac{\sqrt{\alpha} }{2\sqrt{1-\alpha}} + o\left(\frac{1}{\sqrt{n}}\right)
\end{equation}

In particular, the second cumulant is by definition the mean-square displacement of the TP. Although we can see that the transient depends on the tumbling probability, the MSDs display a $\sqrt{n}$ long-time scaling characteristic of the original single file process that is in particular independent of $\alpha$. 

\subsection{Full distribution of positions}

The simple fact that to leading order in $\rho_0$ cumulants of the same parity are equal tells us that distribution associated to these cumulants is a Skellam distribution \citep{skellam-jrss-1946}. In this case, we know that the full distribution function $\mathcal{P}_n(X)$ for any time $n$ is given by 

\begin{equation}
\mathcal{P}_n(X) \underrel{\rho_0 \to 0}{\sim} \exp\left( -\kappa_{\mathrm{even}}(n) \right) \left( \frac{\kappa_{\mathrm{even}}(n) +\kappa_{\mathrm{odd}}(n) }{\kappa_{\mathrm{even}}(n) - \kappa_{\mathrm{odd}}(n) } \right)^{X/2} I_X \left( \sqrt{\kappa_{\mathrm{even}}(n)^2 -\kappa_{\mathrm{odd}}(n)^2}\right),
\end{equation}

where $I_X$ is a modified Bessel function of the first kind \citep{abramowitz-book-1972}. We show in the main text a perfect agreement between this distribution and the results of our numerical simulations.

Here, in the long time limit, the distribution reads

\begin{equation}
\mathcal{P}_n(X) \underrel{\rho_0 \to 0}{\sim} \exp\left( -\rho_0 \sqrt{2n/\pi} \right)  I_X \left( \rho_0\sqrt{2n/\pi}\right).
\label{Seq:distributionsBiasedRT}
\end{equation}

\end{document}